\documentclass[12pt,a4paper]{article}
\usepackage{graphicx}
\usepackage{times}
\title{\bf Non-thermal radiation from a runaway massive star}
\author{Gustavo E. Romero$^{1,2}$, Paula Benaglia$^{1,2}$, 
Cintia S. Peri$^{1,2}$,\\
Josep Mart\'i$^{3}$, Anabella T. Araudo$^{1,2}$\\
\vspace{1cm}\\
\normalsize $^1$ Instituto Argentino de Radioastronom\'ia, CCT-La Plata, 
CONICET,\\
\normalsize  C.C.5, (1894) Villa Elisa, Buenos Aires, Argentina\\ 
\normalsize $^2$ Facultad de Ciencias Astron\'omicas y Geof\'isicas UNLP,\\
\normalsize Paseo del Bosque S/N, B1900FWA La Plata, Argentina\\
\normalsize $^3$ Departamento de F\'isica (EPS), Universidad de Ja\'en, 
\normalsize Campus Las Lagunillas s/n, 23071 Ja\'en, Spain}
\date{\mbox{}}
\begin{document}
\maketitle
\pagestyle{empty}
%
%
\def\bull{\vrule height .9ex width .8ex depth -.1ex}
\makeatletter
\def\ps@plain{\let\@mkboth\gobbletwo
\def\@oddhead{}\def\@oddfoot{\hfil\tiny\bull\quad
``The multi-wavelength view of hot, massive stars''; 39$^{\rm th}$ Li\`ege Int.\ Astroph.\ Coll., 12-16 July 2010 \quad\bull}%
\def\@evenhead{}\let\@evenfoot\@oddfoot}
\makeatother
%
%
\def\beginrefer{\section*{References}%
\begin{quotation}\mbox{}\par}
\def\refer#1\par{{\setlength{\parindent}{-\leftmargin}\indent#1\par}}
\def\endrefer{\end{quotation}}
%
%
{\noindent\small{\bf Abstract:}  We present a study of the radio
  emission from a massive runaway star. The star forms a bow shock
  that is clearly observed in the infrared. We have performed VLA
  observations under the assumption that the reverse shock in the
  stellar wind might accelerate charged particles up to relativistic
  energies. Non-thermal radio emission of synchrotron origin has been
  detected, confirming the hypothesis. We have then modeled the system
  and we predict a spectral energy distribution that extends up to
  $\gamma$-rays. Under some simplifying assumptions, we find that the
  intensity at high energies is too low to be detected by current
  instruments, but the future Cherenkov Telescope Array might detect
  the source.}

\section{Introduction}
Runaway OB stars (Gies and Bolton 1986) can produce so-called stellar
bow shocks on the surrounding interstellar medium. Bow shocks develop
as arc-shaped structures, with bows pointing in the same direction of
the stellar velocity, while the star moves supersonically through the
interstellar medium (ISM). The stellar and shock-excited radiation
heat the dust and gas swept by the bow shock. The dust, in turn,
re-radiates the energy as mid to far IR flux.

\begin{figure}[h]
\centering
\includegraphics[width=6cm,angle=-90]{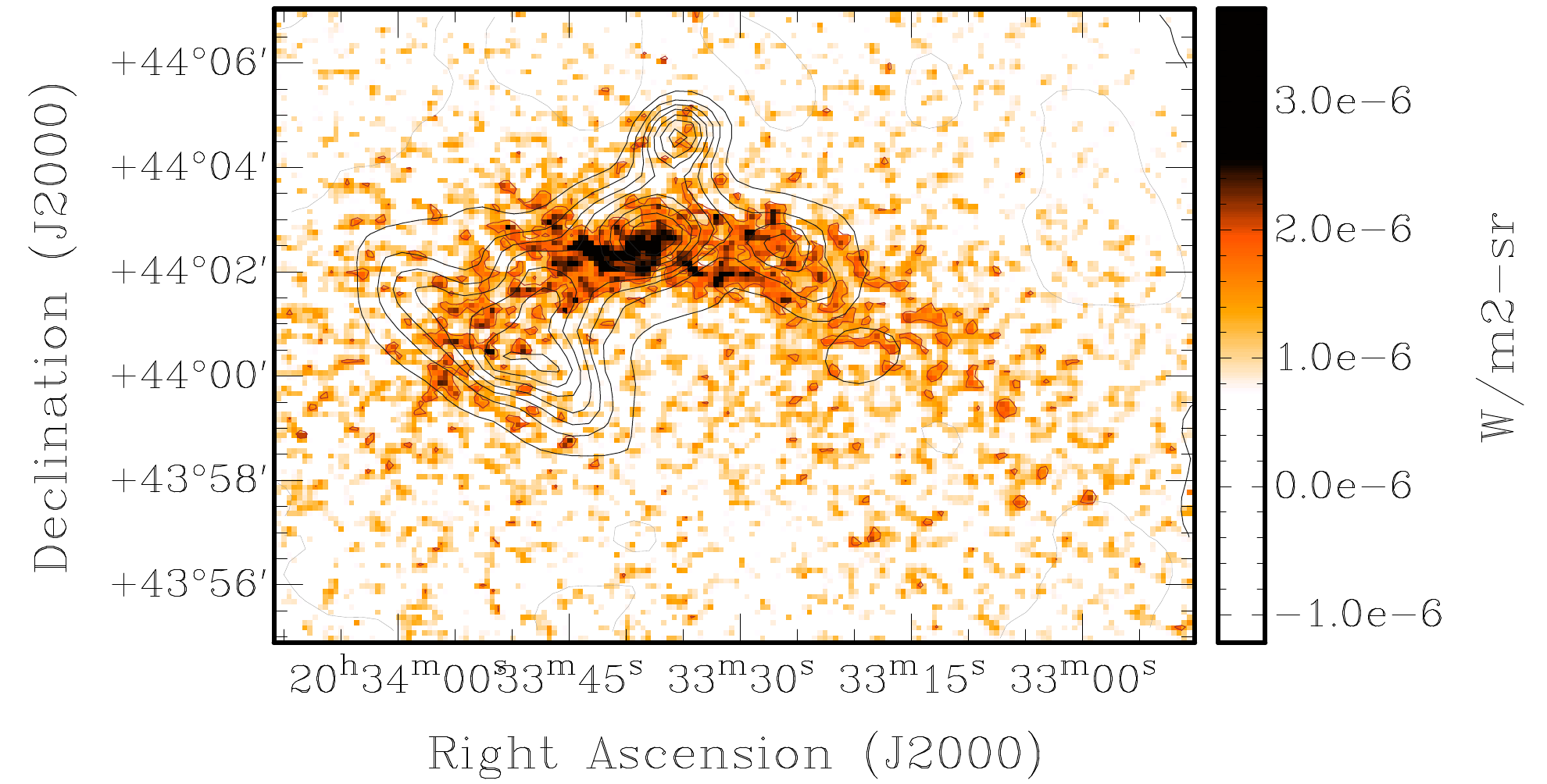}
\caption{MSX-D band image (color scale) superposed to 1.4 GHz-NVSS contours. 
Levels are: $-$2, 2 (2$\sigma$), 5, 8, 11, 15, 19, 24, 29, 50, 70, and 90 
mJy beam$^{-1}$.\label{fig_1}}
\end{figure}

Van Buren \& McCray (1988) looked for bow-shaped features near
high-velocity O stars and found an IR candidate close to the O
supergiant BD+43$^{\circ}3654$ ($\alpha,\delta$[J2000] = $20^{\rm
  h}33^{\rm m}36.077^{\rm s}, +43^{\circ} 59' 07.40''$; $l, b =
82.41^\circ, +2.33^\circ$). Comer\'on \& Pasquali (2007) related the
star BD+43$^{\circ}$3654 to a bow shock detected with the Midcourse
Space eXperiment (MSX) at D and E bands. Also, data from the
NRAO-VLA\footnote{National Radio Astronomy Observatory - Very Large
  Array (http://www.vla.nrao.edu/).} NVSS Survey (Condon et al. 1998)
revealed a coma-shaped source of $\sim$ 7 arcmin spatially coincident
with the MSX feature (Figure \ref{fig_1}).

A radio study of the bow shock can shed light on the physical
processes that give rise to high-energy emission from a stellar
source, regardless of the history of the runaway star. The shock can
accelerate particles up to relativistic energies by Fermi
mechanism. Energetic electrons will cool through synchrotron
radiation, producing a non-thermal radio source. We carried out radio
observations at two frequencies to study the nature of the emission
from the bow shock of BD+43$^{\circ}$3654.

\section{Observations and results}
\begin{figure}[h]
\centering
\label{fig_2}
\includegraphics[width=5.3cm,angle=-90]{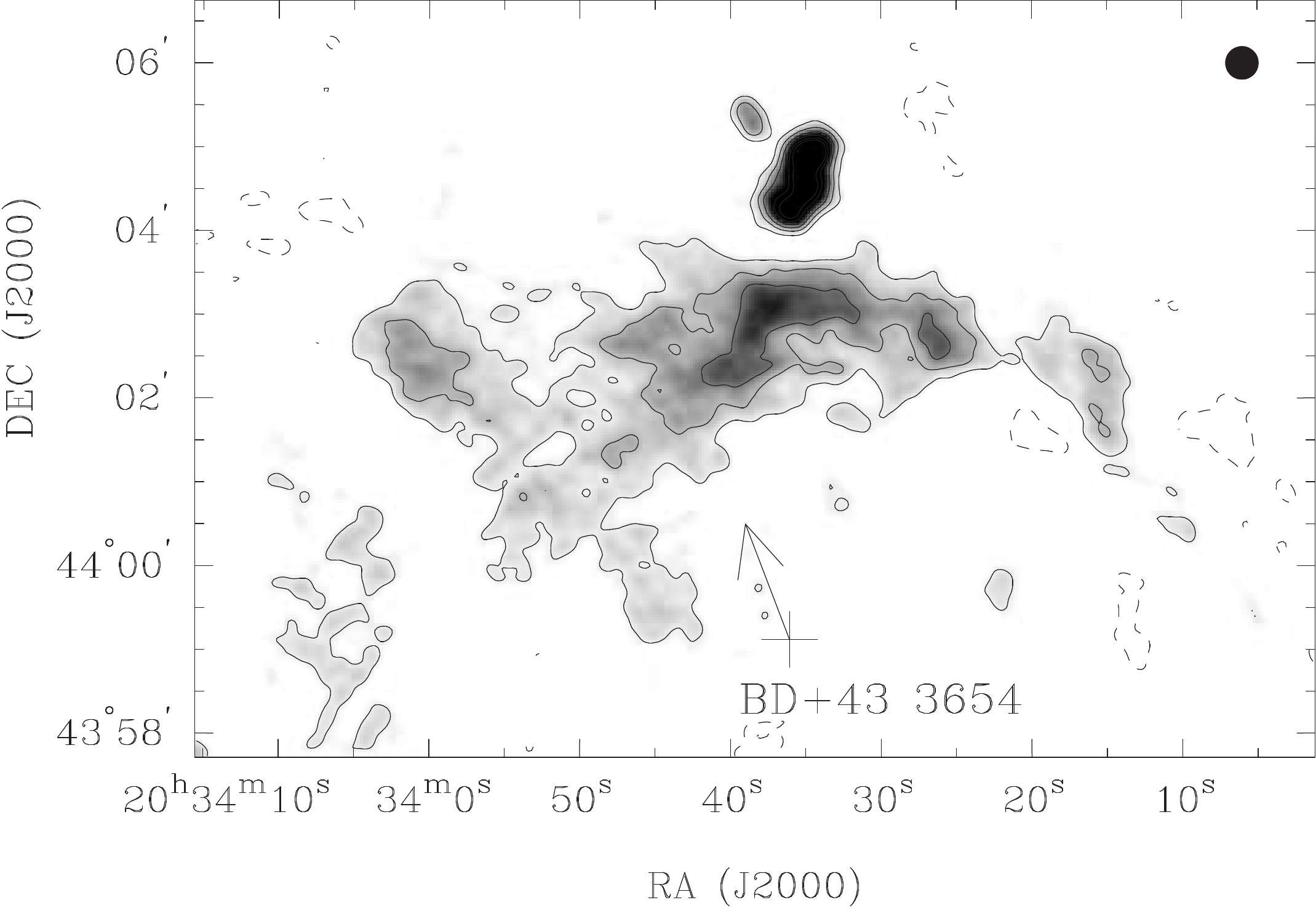}\hspace{1cm}
\includegraphics[width=5.3cm,angle=-90]{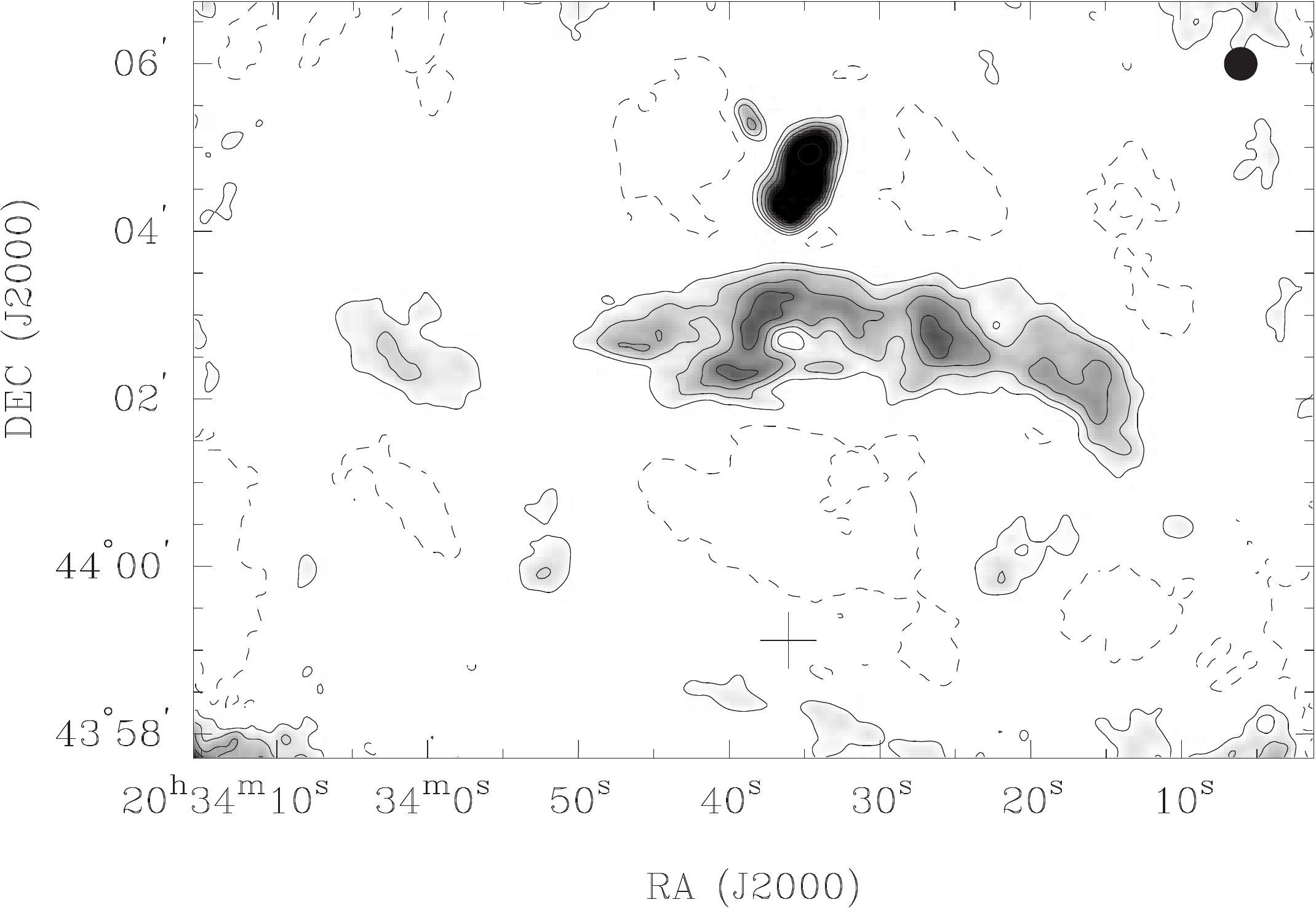}
\caption{Continuum emission at 1.42 GHz (left), and at 4.86 GHz (right). 
Contour levels are $-3$, 3, 6, 10, 15, 20, 25, and 60 times the rms of 0.3 
and 0.2 mJy beam$^{-1}$. BD+43$^{\circ}\,3654$ is marked with a cross. The 
arrow represents the velocity vector of the star, derived from proper motions 
corrected for local motion of the surrounding ISM (see text). Synthesized 
beams of $12''\times 12''$ are shown in the top right corners.}
\end{figure}

Our continuum observations were carried out with the Very Lare Array
(VLA) at 1.42 GHz (C config.) and at 4.86 GHz (D config.). Figure 2
presents the resulting images after primary beam correction,
re-gridded with the same synthesized beam of 12''. There is emission
at both frequencies along the extension of the MSX source. The
hypothesis of a physical association between the star and the radio/IR
features is supported by the very good agreement of the residual
proper motion of the star and the direction from the star to the apsis
of the bow shock (Fig. \ref{fig_2}).
We used the continuum images at 1.42 and 4.86 GHz to build a spectral
index distribution map. We only considered input pixels with a
signal-to-noise ratio $\geq$ 4. Besides, the spectral index map was
masked for a signal-to-noise ratio $\geq$ 10. Figures 3 and 4 show the
spectral index distribution and corresponding noise maps.
\begin{figure}[h]
\centering
\label{fig_4}
\includegraphics[width=6cm,angle=-90]{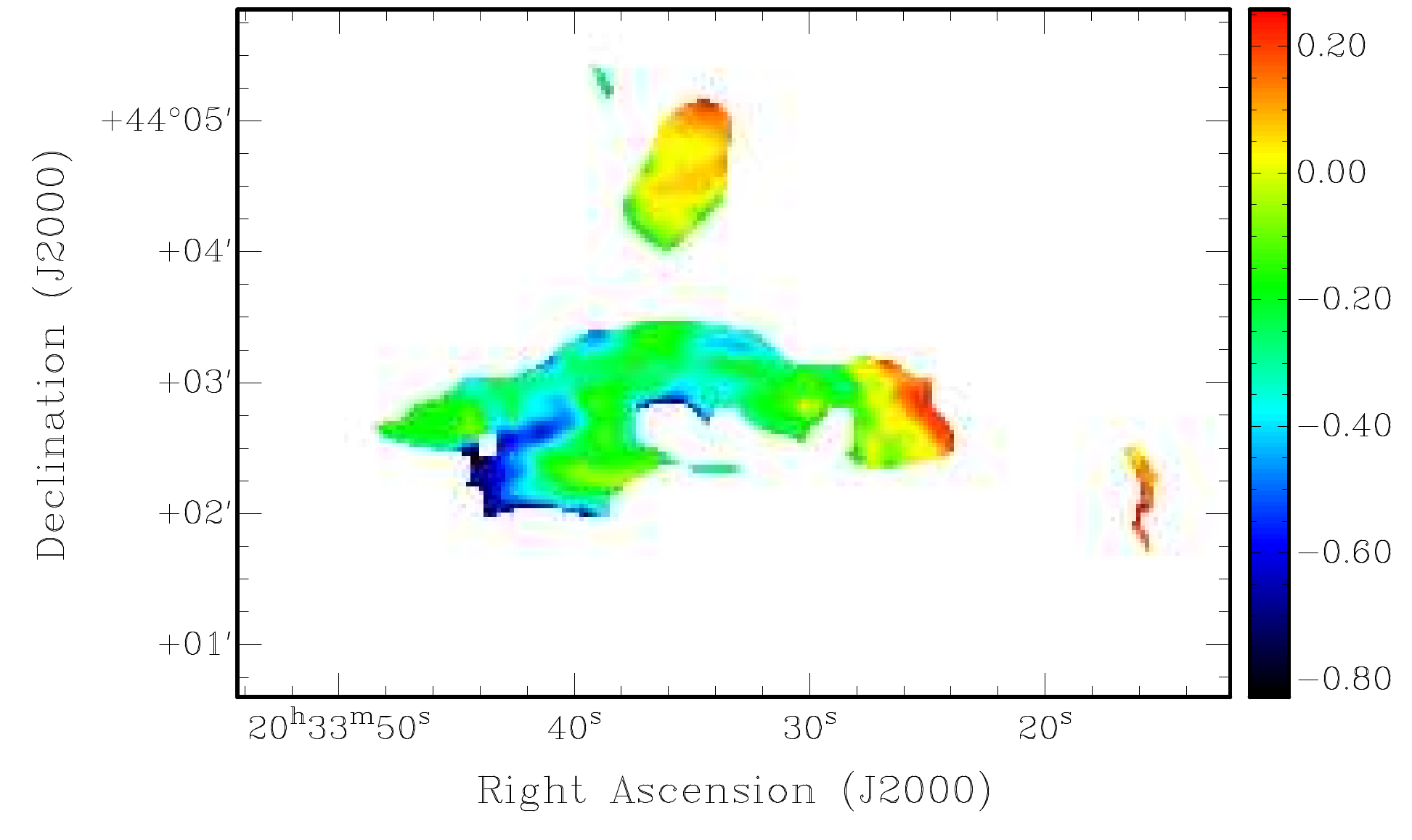}
\caption{Spectral index distribution.}
\end{figure}
\begin{figure}[h]
\centering
\label{fig_5}
\includegraphics[width=6cm,angle=-90]{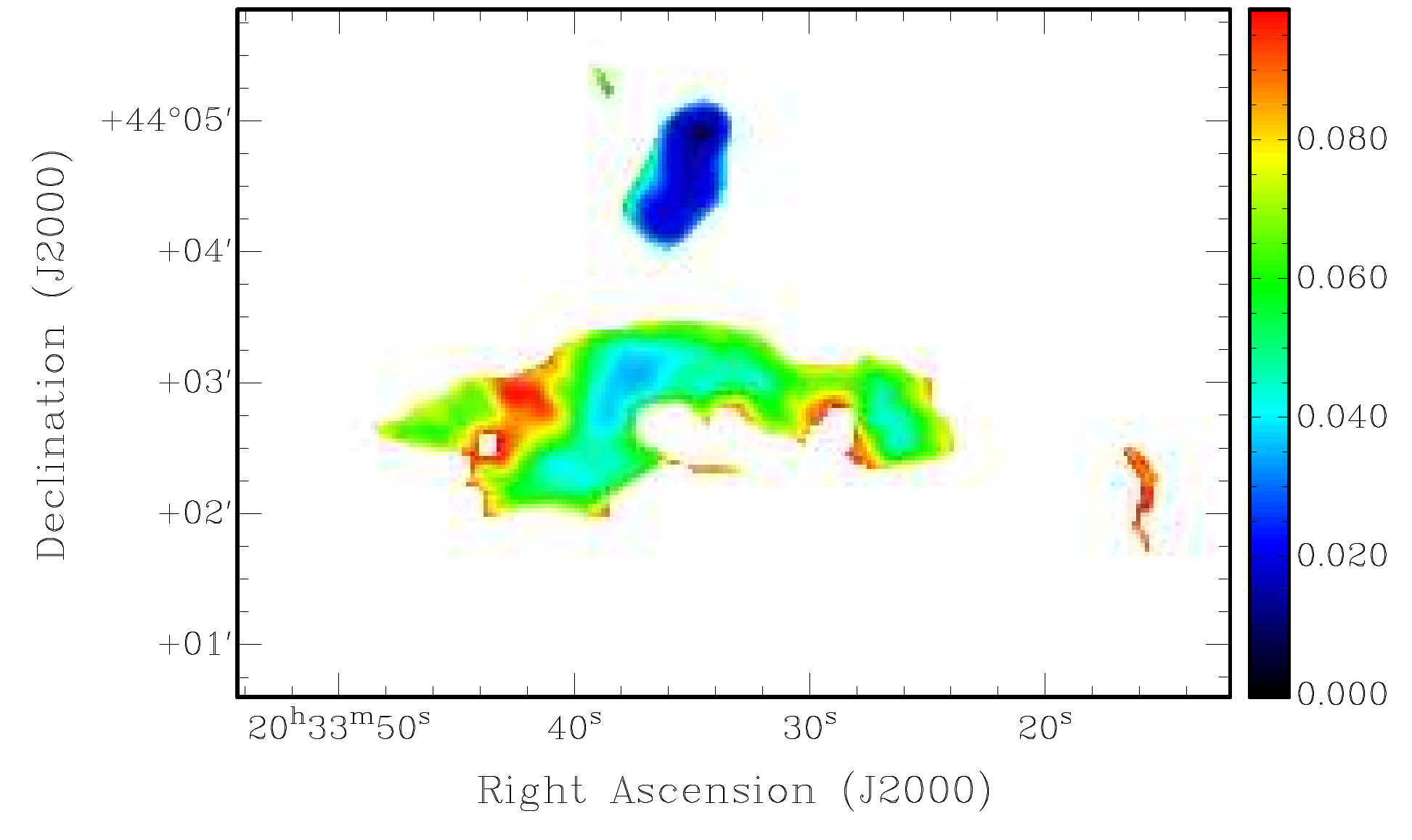}
\caption{Spectral index error distribution.}
\end{figure}
\section{Bow-shock emission}
Most of the area shows a source of non-thermal radiation with index 
$<\alpha>=-0.5$ ($S_{\nu} \propto \nu^{\alpha}$), as obtained from the VLA 
data. We adopted a distance to the bow-shock of 1.4 kpc (Hanson 2003). The 
distance from the star to the bow-shock is $R=5'$, or 2 pc. The volume 
occupied by the bow-shock is $\sim 4.6\, {\rm pc}^3$. We took a particle 
density of the ISM in the bow shock region of 100 cm$^{-3}$ (see Kobulnicky 
et al. 2010). 

The non-thermal radiation is expected from synchrotron emission
generated by relativistic electrons accelerated either at the forward
shock in the ISM or in the reverse shock in the stellar wind. We
estimated the particle energy distribution (n) using the observed flux
density and spectral slope, and assumed equipartition between magnetic
and relativistic particles energy density. 

We considered that the energy density of relativistic particles has
three contributions: 
\begin{equation}
	u=u_{e_1}+u_p+u_{e_2}=\int E_{e_1} n_{e_1}(E_{e_1})
        dE_{e_1}+\int E_{p} n_{p}(E_{p}) dE_{p}+\int E_{e_2}
        n_{e_2}(E_{e_2}) dE_{e_2}, \nonumber
\end{equation}
where $e_1$, $p$, and $e_2$ stand for relativistic primary electrons,
protons, and secondary electron-positron pairs (i.e. pairs coming from
charged pion decays), respectively. The relation between primary
electrons and protons energy density is $u_{p}=a u_{e_1}$, with
$a\geq0$. Three cases were considered: $a=0$ (just electrons), $a=1$
(equal energy density in both species), and $a=100$ (proton-dominated
case, as observed in the galactic cosmic rays). The magnetic field
resulted $B\sim 5 \times 10^{-5}$ G.

The maximum energy of the particles was determined by balancing energy
gains and losses. The loss mechanisms considered were {\sl (i)}
synchrotron radiation, {\sl (ii)} relativistic Bremsstrahlung, {\sl
  (iii)} particle escape from the radiation region due to convection
by the stellar wind, and {\sl (iv)} inverse Compton (IC) scattering of
IR, stellar and cosmic microwave background photons. In the case of
protons, the only relevant losses are proton-proton ($pp$) inelastic
collisions and convective escape. Diffusion is negligible in
comparison to convection in this situation (the respective timescales are
$t_{\rm conv} \sim 6\times10^6$~s and 
$t_{\rm diff} \sim 10^{13}/(E/{\rm erg})$~s). Both primary electrons and
protons reach energies up to $\sim 10^{13}$ eV, which is imposed by
non-radiative losses, except for $a=100$, where synchrotron losses
dominate for electrons. Figure 5 shows the losses for electrons and
protons in the case $a=1$.

\begin{figure}[h]
\centering
\includegraphics[width=8cm]{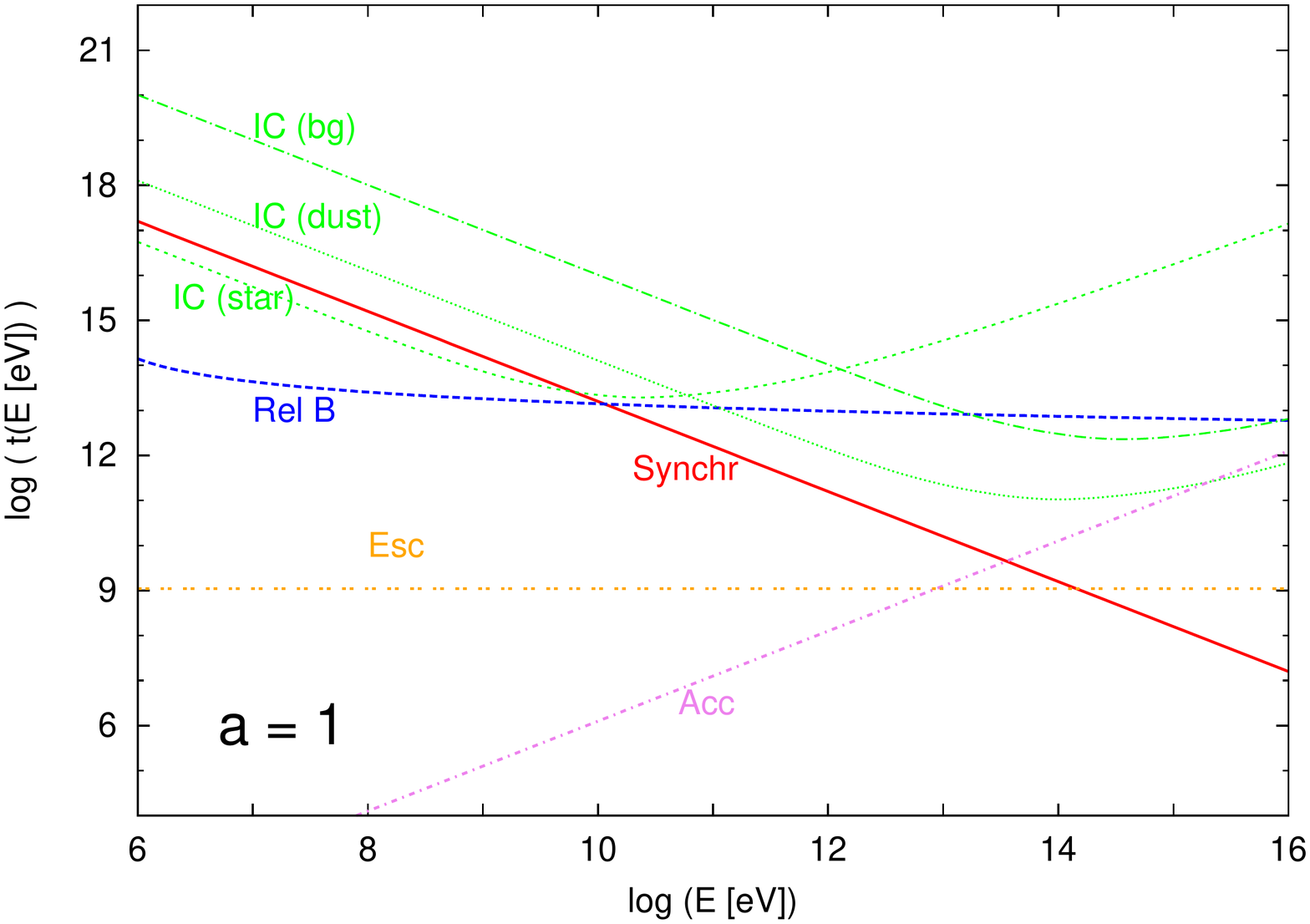}
\includegraphics[width=8cm]{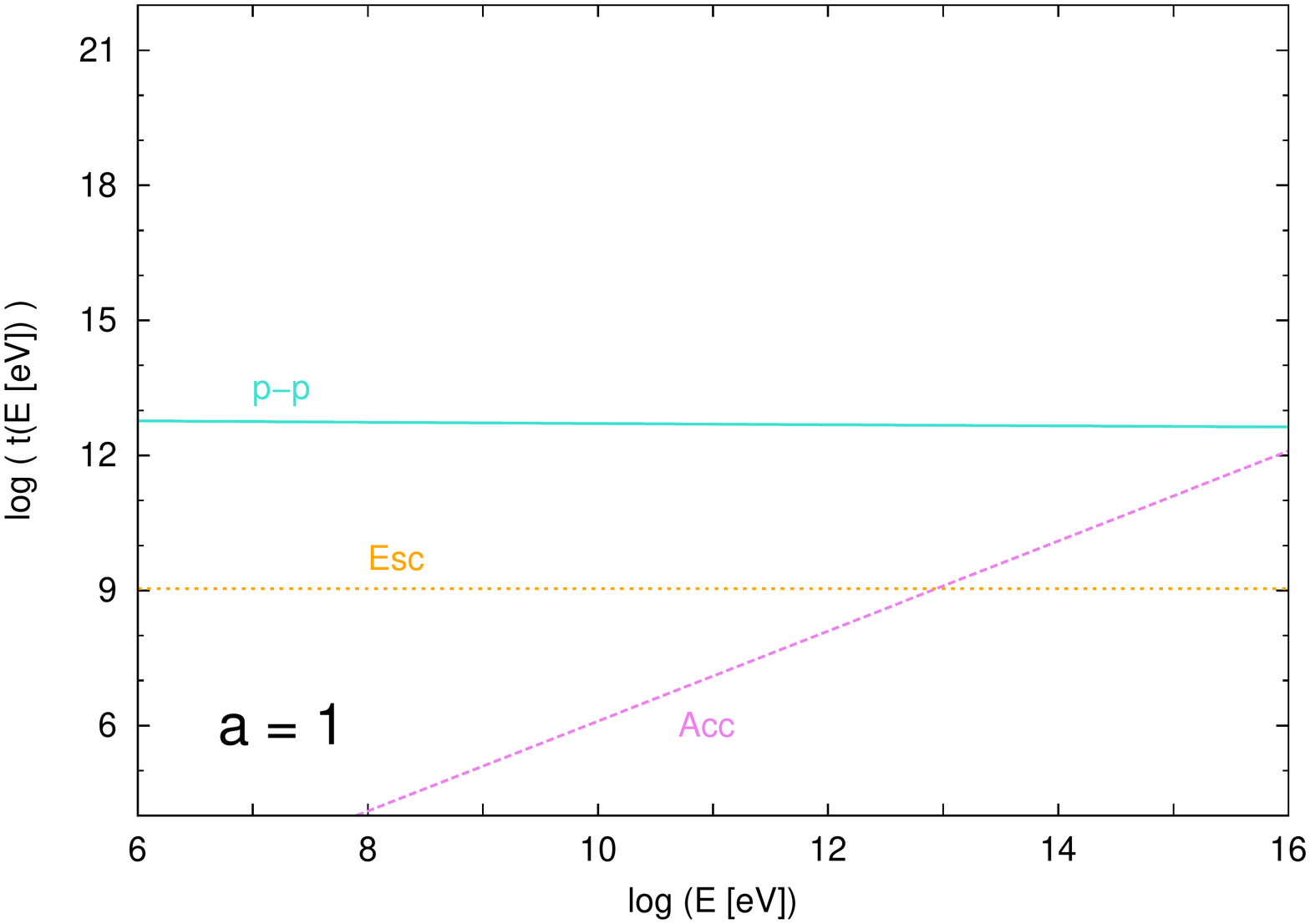}
\label{fig_6}
\caption{Left: acceleration (`Acc'), escape (`Esc'), and cooling
  times for electrons, due to synchrotron radiation (`Synchr'), IC
  scattering of dust photons (`IC (dust)'), stellar photons (`IC
  (star)'), and backgroud photons (`IC (bg)'). Cooling time for
  relativistic Bremsstrahlung radiation is indicated as `Rel  B'. The
  figure is for the case with equal energy density in electrons and
  protons ($a=1$, see text). Right: acceleration, escape, and cooling 
time for protons due to the $pp$ radiation
  process ('p-p').} 
\end{figure}

\section{Discussion and perspectives}
The presence of highly relativistic particles in a dense medium with
high photon density can result in the efficient generation of
$\gamma$-rays. The corresponding $\gamma$-ray emissivity can be
calculated using the delta-functional approximation (e.g. Aharonian \&
Atoyan 2000, Kelner et al. 2006).

In Fig. 6 we show the spectral energy distribution obtained for the
case $a=1$, with all contributions included (synchrotron self-Compton
losses are negligible). The total luminosity from $pp$ interactions is
similar to that obtained from relativistic Bremsstrahlung of
electrons. The IC up-scattering of IR photons is the major
contribution at high energies, with a peak around 100 GeV. The
detectability of the source by the Fermi LAT\footnote{Fermi Large Area
  Telescope (http://www-glast.stanford.edu/).} $\gamma$-ray
observatory will depend on the actual particle density and the
contribution related to the secondary electrons at large $a$. The $pp$
contribution extends well into the TeV regime, but its weaker and will
be difficult to detect with the current ground-based Cherenkov
telescope arrays.

For the case $a=100$, the relativistic particle contain is
proton-dominated and $\gamma$-rays from $pp$ process dominate the high
energy spectrum. The CTA\footnote{Cherenkov Telescope Array
  (http://www.cta-observatory.org/).} North observatory might detect
the source yielding information on the high enery cutoff.

Observations of the spectral slope at high energies can be used to
identified the proton content through the luminosity level, and the
proton spectral index. Radio polarization data will provide additional
information of $B$. X-ray observations will allow to determine the
cutoff of the synchrotron spectrum, directly related to the maximum
energy of the electrons\footnote{Notice that the situation is quite 
different
from that of colliding winds, where the particle acceleration occurs in a 
region of high photon density, with dominance of IC losses.}. 
This, in turn, would yield valuable
information on the actual value of $B$ and the correctness of the
equipartition hypothesis.

\begin{figure}[h]
\centering
\label{fig_8}
\includegraphics[width=12cm]{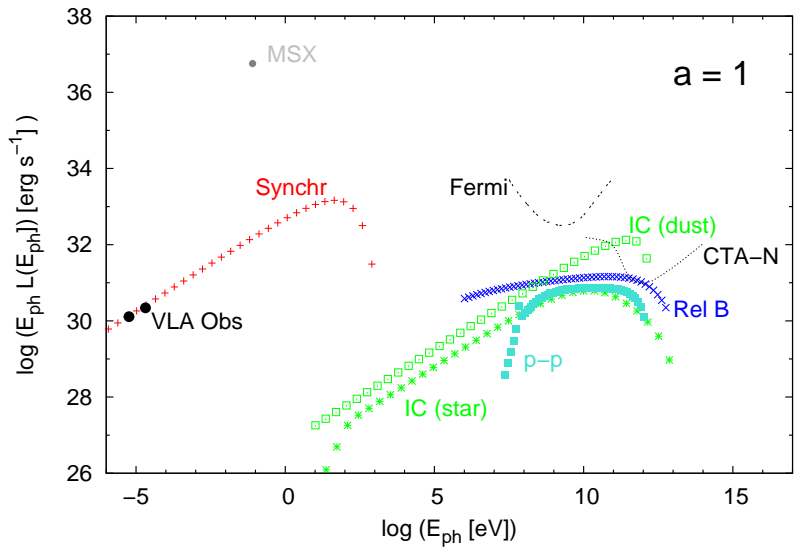}
\caption{Spectral energy distribution for the case $a=1$. Acronyms as in 
Figure 5. Measured radio fluxes from VLA observations (`VLA Obs') and MSX 
luminosity at D band are also represented. The contribution from secondary 
pairs is negligible in this case, so is not shown here.}
\end{figure}

\section*{Acknowledgements}
This work was supported by MinCyT - ANPCyT 
(PICT-2007-00848) 
and by CONICET 
(project ID 11220090100078). JM and GER 
acknowledge support by grant AYA2007-68034-C03-01 and -02 from the Spanish 
government and FEDER funds 
and Plan Andaluz de 
Investigaci\'on, Desarrollo e Innovaci\'on of Junta de Andaluc\'{\i}a as 
research group FQM-322 and excellence fund FQM-5418.
%
%
\footnotesize
\beginrefer
\refer Aharonian, F. A., Atoyan, A. M. 2000, A\&A, 362, 937

\refer Comer\'on, F., Pasquali, A. 2007, A\&A, 467, L23

\refer Condon, K. K, Cotton, W. D., Greisen, E. W., et al. 1998, AJ, 115, 1693

\refer Gies, D., Bolton, C. T. 1986, ApJS, 61, 419

\refer Hanson M. M. 2003, ApJ 97, 957-969

\refer Kelner, S. R., Aharonian, F. A., Bugayov, V. V. 2006, Phys. Rev. D, 
74, 4081

\refer Kobulnicky, H. A., Gilbert, I. J., Kiminki, D. C. 2010, ApJ, 710, 529

\refer Van Buren, D., McCray, R. 1988, ApJ, 329, L93

\endrefer           

\end{document}